# Sensor Fusion for Track Geometry Monitoring: Integrating On-Board Condition Monitoring and Degradation Models via Kalman Filtering

Huy Truong-Ba[1,*], Jacky Chin[2], Michael E. Cholette[1], Pietro Borghesani[2]

[1] *Science and Engineering Faculty, Queensland University of Technology, Brisbane Queensland, Australia*

[2] *School of Mechanical and Manufacturing Engineering, UNSW Sydney, Australia*

**Corresponding author:* h.truongba@qut.edu.au

## Abstract

Track geometry monitoring is essential for maintaining the safety and efficiency of railway operations. While Track Recording Cars (TRCs) provide accurate measurements of track geometry indicators, their limited availability and high operational costs restrict frequent monitoring across large rail networks. Recent advancements in on-board sensor systems installed on in-service trains offer a cost-effective alternative by enabling high-frequency, albeit less accurate, data collection. This study proposes a method to enhance the reliability of track geometry predictions by integrating low-accuracy sensor vibration signals with degradation models through a Kalman filter framework. An experimental campaign using a low-cost sensor system mounted on a TRC evaluates the proposed approach. The results demonstrate that incorporating frequent sensor data significantly reduces prediction uncertainty, even when the data is noisy. The study also investigates how the frequency of data recording influences the size of the credible prediction interval, providing guidance on the optimal deployment of on-board sensors for effective track monitoring and maintenance planning.

## 1 Introduction

Track geometry is a critical factor in ensuring the safety and comfort of train operations [1–4]. Rail operators must continuously monitor various track geometry indicators to identify deviations from acceptable operating limits and to carry out timely maintenance interventions. Among these maintenance actions, tamping is the most widely adopted method for correcting track geometry.

Tamping involves adjusting the position of the track to restore its geometry, and it can be performed either manually or with specialized mechanized tamping machines [5]. The decision to undertake tamping is typically based on the condition of the track geometry indicators as recorded by the Track Recording Car (TRC) [6,7].

The TRC plays a vital role in monitoring and maintaining track geometry by providing precise measurements of key indicators. However, the availability of TRCs is often limited, particularly for rail operators managing extensive rail networks where TRCs are in high demand. Maintaining smooth train operations across large networks requires an adequate number of TRCs, but procuring and maintaining these vehicles entails significant capital investment. As a result, there is growing interest in alternative monitoring systems that are both cost-effective and capable of providing frequent updates on track geometry, albeit with slightly lower accuracy. Such systems can serve as a complementary tool, guiding decision-makers in identifying tracks that require more detailed TRC inspection [8,9]. This study focuses on enhancing the effectiveness of track geometry monitoring by exploring this alternative approach.

Predicting track geometry indicators is an essential approach for optimizing TRC deployment and planning maintenance activities. Accurate predictions enable rail operators to prioritize tracks for inspection and intervention, thus ensuring that maintenance resources are allocated efficiently. Over the years, numerous studies have been conducted to develop methodologies for estimating track geometry degradation. These methodologies can be broadly classified into two main categories: mechanical models [6,10] and statistical models [11–14]. Statistical or data-driven degradation models are more favorable due to their capability of including the uncertain factors that mechanical models cannot fully consider [15]. Among statistical techniques, Gamma processes [16] and Wiener processes [4,11,17] are the most commonly used for formulating track geometry degradation models, in which the Wiener processes are preferable because they are well-suited for capturing the stochastic nature of track geometry changes over time, allowing for more realistic predictions.

Previous research has predominantly focused on modeling the degradation of individual track geometry indicators or composite indicators, such as track quality indices [13,18]. However, there has been limited exploration of models that consider multiple track geometry indicators simultaneously. One notable exception is the work by Mercier et al. [16], who developed a bivariate Gamma process model to account for both longitudinal and transversal leveling indicators in track geometry degradation. Building on this foundation, our previous work [19] investigated the correlations among various track geometry indicators and introduced a multivariate Wiener model to represent the degradation process with multiple indicators. This approach offers a more comprehensive understanding of track geometry degradation, enabling more accurate and robust predictions that can inform maintenance planning and decision-making.

The most widely adopted alternative to TRCs for monitoring track geometry condition is the use of in-service train systems, in which various sensors are installed on operational trains [8,9,20–22]. Existing studies have primarily focused on the development of sensing technologies, including vibration-based sensors [8,23,24], topographic profile devices [25], or vision sensors [26,27]. To improve the accuracy of track geometry estimation obtained from on-board sensor systems, filtering approaches—particularly those based on the Kalman filter family—have been extensively investigated. Munoz et al [28,29] and De Rosa et al [30] applied Kalman filtering to estimate lateral and vertical track irregularities, while unscented and nonlinear extended Kalman filters were employed in [31] to enhance track geometry measurements using a track trolley. More recently, a Bayesian Kalman filter was proposed in [32] to quantify estimation uncertainty arising from vehicle–bridge interaction effects. Despite these advances, the accuracy of on-board sensor systems remains strongly dependent on sensor cost and quality [33,34], and differences in operating conditions between in-service trains and TRCs—particularly speed variability—can further degrade estimation performance. These limitations raise a key question regarding how to effectively exploit the more frequently collected but less accurate data from low-cost on-board sensors for reliable track geometry condition estimation.

Degradation models play a critical role in enhancing TRC planning by enabling the prioritization of tracks that exhibit rapid degradation for timely inspections. However, these models have inherent limitations, particularly the growth of credible intervals when predictions extend far beyond the most recent observations obtained from the latest TRC run. This increasing uncertainty in track geometry degradation forecasts highlights the need for strategies to reduce predictive uncertainty. One effective approach is to increase the frequency of track geometry measurements. While increasing the number of TRC inspections is often economically infeasible, the deployment of sensor systems on in-service trains—referred to as on-board sensor systems—offers a viable alternative [8]. Achieving accuracy comparable to that of TRCs using on-board systems would require high-grade sensors and associated technologies, which undermines their economic advantage. To address this trade-off, this study proposes a cost-effective hybrid framework that integrates infrequent high-accuracy TRC measurements with frequent low-cost on-board sensor data. Although the measurements obtained from on-board systems are not sufficiently accurate to fully replace TRCs, their high temporal resolution can significantly enhance track geometry estimation and reduce the required number of TRC runs.

It is important to note that the filtering approaches cited above [28–32] addressed a fundamentally different estimation problem: they seek to reconstruct instantaneous track irregularity profiles from vehicle dynamic response during a single train pass. The present work, by contrast, formulates a long-horizon state estimation problem in which the system state evolves according to a stochastic degradation process over weeks to months, and is periodically updated with noisy on-board sensor observations. In doing so, this work bridges two research streams that have so far developed largely in parallel: degradation modelling of track geometry indicators based on TRC measurements, and on-board sensor-based geometry estimation. The fusion of sparse high-accuracy inspection data with frequent low-cost sensor measurements for degradation tracking has, to the authors' knowledge, not been previously investigated in the track geometry monitoring literature. This formulation also enables a systematic analysis of how measurement frequency affects predictive uncertainty, which is a practical question directly relevant to the deployment of on-board sensor systems.

This study aims to develop a robust method for improving the reliability of track geometry predictions by leveraging signals from sensors installed on in-service trains. The effectiveness of this method is evaluated through an experimental campaign that incorporates a low-cost sensor system integrated into a real TRC. The proposed approach seeks to enhance the credibility of track geometry predictions, even when based on frequently recorded but inherently noisy signals, through the application of a Kalman filter. This filter allows for the integration of real-time data and provides a means to systematically reduce prediction uncertainty. Furthermore, the study analyzes the relationship between the frequency of recording data and the resulting reduction in the credibility zone, yielding valuable insights into the optimal number of sensor systems required on service trains to achieve effective monitoring.

The remainder of this paper is organized as follows: Section 2 describes the fusion methodology of TRC data and on-board measurement via a Kalman filter approach. Section 3 outlines the experimental campaign, detailing the low-cost sensor system installed on the TRC and its operational parameters. Section 4 examines the implications of recording frequency for prediction credibility, highlighting its significance in enhancing the reliability of track geometry assessments. Finally, the concluding section summarizes the key contributions of this study and explores potential avenues for future research, emphasizing the importance of continued innovation in track geometry monitoring and prediction methodologies.

# 2 Fusion of Track Recording Car and on-board measurements via State Estimation

## 2.1 Overall Concept

The TRC provides highly accurate measurements of track geometry, but its high cost and limited availability make it difficult to obtain geometry indicators within the required time frame. To address this challenge, it is useful to deploy lower-cost sensor systems on service trains. Although these systems are less accurate, they are significantly cheaper and can collect track data at a much higher frequency. With the implementation of on-board sensor systems, there are now two sources of track geometry data with a different balance of accuracy and frequency: the TRC is accurate but infrequent, while the on-board sensor system is more frequent (owing to its installation on a number of trains) but less accurate data. This scenario motivates the development of a method to exploit the strength of both approaches for the collaborative estimation of track geometry.

The research problem in this study is described as follows: two time-series datasets are available for estimating the state of the system. One dataset is more frequent but less certain (i.e. "noisier"), while the other is less frequent but more reliable. In this study, we consider this latter measurement to represent the system's true state. The more frequent dataset can be used to estimate the state of the system, which is periodically recorded by the less frequent but more accurate process, which reveals the true system state (Figure 1). Since the state recording process provides accurate system states, the estimation is re-started each time the system state is observed by the accurate inspection. This collaborative estimation of system states based on these datasets can be approached as an (occasionally re-started) filtering problem [35].

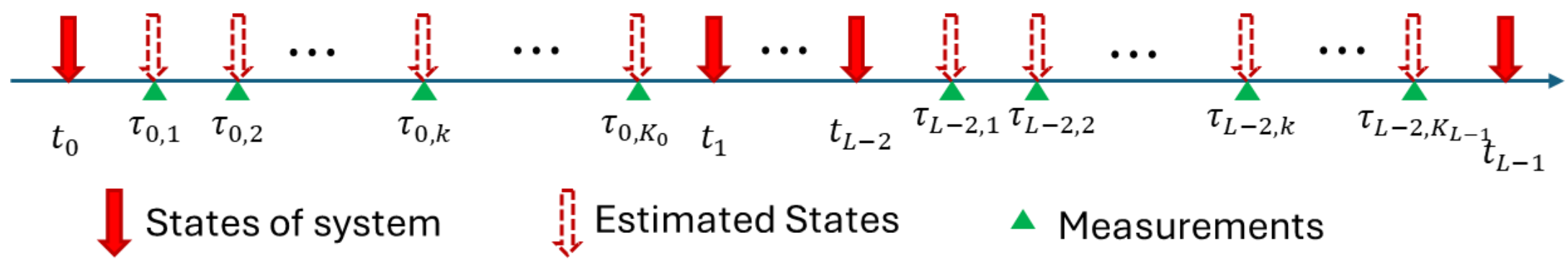

Figure 1: Illustration of collaborative estimation problem.

A significant body of literature has explored and applied various filtering methods to address system state estimation problems. The ubiquitous Kalman filter is the standard approach for handling linear filtering problems [36]. For nonlinear problems, several modified versions of the Kalman filter have been successfully employed in both research and practical applications, including the Extended Kalman filter [36], Robust Kalman filter [37], Unscented Kalman filter [38], Cubature Kalman filter [39], or Probabilistic Kalman filter [40]. In addition to these, other filtering methods, such as the predictive filter [41], robust filter [42], and particle filter [43], are frequently used for tackling nonlinear problems. Among these, particle filtering methods, which are based on sequential Monte Carlo methods and Bayesian inference, have gained increasing popularity for addressing complex issues, particularly in the presence of non-Gaussian noise [44].

The general filtering problem is described as follows. Given a time series $t_k, k = 0,1, \ldots$, let $\boldsymbol{X}_k \in \mathbb{R}^n$ be the sequence of random (unobserved) system states and $\boldsymbol{Y}_k \in \mathbb{R}^m$ be the associated observations. The filtering problem is to estimate sequentially the hidden state $\boldsymbol{X}_k$ given the observations $\boldsymbol{y}_0, \boldsymbol{y}_1, \ldots, \boldsymbol{y}_k$. The general transition equation and observation equation have the forms:

$$\boldsymbol{X}_k | (\boldsymbol{X}_{k-1} = \boldsymbol{x}_{k-1}) \sim f(\boldsymbol{x}_{k-1} | \boldsymbol{\theta})$$

$$\boldsymbol{Y}_k | (\boldsymbol{X}_k = \boldsymbol{x}_k) \sim g(\boldsymbol{x}_k | \boldsymbol{\theta})$$

where $\boldsymbol{\theta}$ denotes the model parameters, $f(\cdot)$ is the transition function/distribution of system states and $g(\cdot)$ is the distribution of the observations given the state. If both functions are Gaussian, the well-known Kalman filter is the optimal filter.

## 2.2 Prediction and estimation of track geometry with on-board signal

In this study, the horizon of $L$ accurate recording times of TRC records is denoted as $t_\ell, \ell = 0,1, \ldots L-1$ with $t_0 < t_1 < \cdots < t_{L-1}$ which may be unevenly sampled. The on-board signals are collected at time samples $\tau_{\ell k}$ with $t_\ell < \tau_{\ell,k} < t_{\ell+1}$ and $k = 1,2, \ldots, K_\ell$. The sequence of measurement times is therefore $t_0, \tau_{0,1}, \tau_{0,2}, \ldots, \tau_{0,K_0}, t_1, \tau_{1,1}, \tau_{1,2}, \ldots, \tau_{(L-1),K_{L-1}}, t_{L-1}$ (e.g. as in Figure 1). For these time points, the $\ell$ subscript denotes the index of the previous TRC recording time, while $k$ is the index of the on-board measurement times.

Consider a particular segment of track. It is assumed that $n$ of *geometry indicators* are used to represent the condition of the track geometry at any time $t_k$. This set of geometry indicators is denoted as $\boldsymbol{Z}_k = [Z_{q,k}]_{q=1 \ldots n}$. Following the assumption that TRC measurements are equivalent to direct observations of the track geometry, the value of $\boldsymbol{Z}_k$ becomes known when a TRC record is available.

Similarly, a series of $m$ *on-board indices* $\boldsymbol{Y}_k = [Y_{q,k}]_{q=1 \ldots m}$ are obtained by processing the signals measured by the on-board system. In general, $m \neq n$ and in fact the on-board indices in $\boldsymbol{Y}_k$ can differ in nature from the TRC indicators in $\boldsymbol{Z}_k$, as long as there is a sufficient statistical relationship between the two.

It is worth clarifying that, in this study, the capital-letter symbols $\boldsymbol{Z}_k$ and $\boldsymbol{Y}_k$ are used to denote random variables, while their values are represented with the lower-case symbols $\boldsymbol{z}_k = \left[z_{q,k}\right]_{q=1\ldots n}$ and $\boldsymbol{y}_k = \left[y_{q,k}\right]_{q=1\ldots m}$, respectively[1].

The degradation dynamics of the $n$ geometry indicators are described by a multivariate Wiener model adapted from our previous study [19]. In summary, the degradation during the interval $(t_{k-1}, t_k]$ is described by a normal distribution, i.e., the geometry indicators $\boldsymbol{Z}_k$ at $t_k$, conditional on $\mathbf{Z}_{k-1}$, are:

$$\boldsymbol{Z}_k | (\boldsymbol{Z}_{k-1} = \mathbf{z}_{k-1}) \sim N(\boldsymbol{\mu}_k, \mathbf{Q}_k) \tag{1}$$

The evolution of the expected value $\boldsymbol{\mu}_k$ and covariance matrix $\mathbf{Q}_k$ of the geometry indicators, conditional to $\mathbf{z}_{k-1}$, have expressions that depend on the occurrence or absence of a maintenance (i.e. tamping) event during the interval $(t_{k-1}, t_k]$. Therefore, the mean and covariance matrix are:

$$\boldsymbol{\mu}_k = \begin{cases} \mathbf{z}_{k-1} + \boldsymbol{\mu}\Delta t_k & k \notin \mathcal{M} \\ \mathbf{z}^{+} + \dfrac{1}{2}\boldsymbol{\mu}\Delta t_k & k \in \mathcal{M} \end{cases} \quad \text{and} \quad \mathbf{Q}_k = \begin{cases} \boldsymbol{\Sigma}\Delta t_k & k \notin \mathcal{M} \\ \boldsymbol{\Sigma}_{\mathrm{z}^{+}} + \dfrac{\boldsymbol{\Sigma}\Delta t_k}{2} & k \in \mathcal{M} \end{cases} \tag{2}$$

where $\mathcal{M}$ is the set of time intervals in which maintenance is conducted, i.e., $k \in \mathcal{M}$ if maintenance occurs in the interval $(t_{k-1}, t_k]$. The expressions above also assume that maintenance occurs in the middle of the interval at time $(t_{k-1} + t_k)/2$. The variable $\Delta t_k = t_k - t_{k-1}$ represents the duration of the interval $k$, while $\boldsymbol{\mu}$ and $\boldsymbol{\Sigma}$ are the drift and diffusion matrices of selected indicators from degradation model. The additional terms $\boldsymbol{z}^{+}$ and $\Sigma_{\mathrm{z}^{+}}$ are the mean and covariance of the geometry indicators right after tamping.

The on-board indices $\boldsymbol{Y}_k$ are also modelled using a multivariate Normal distribution, conditional on the TRC indicators $\boldsymbol{Z}_k$:

$$\boldsymbol{Y}_k | (\boldsymbol{Z}_k = \mathbf{z}_k) \sim \mathcal{N}(\boldsymbol{m}_k, \mathbf{R}) \tag{3}$$

where $\mathbf{R}$ is the measurement error covariance and $\mathbf{m}_k$ is the indices' mean, which satisfies:

$$\boldsymbol{m}_k = \mathbf{H} \cdot \boldsymbol{z}_k + \mathbf{b} \tag{4}$$

where the matrix $\mathbf{H}$ is the "measurement" sensitivity and $\mathbf{b}$ is a bias term. The $m \times n$ matrix $\mathbf{H}$ and the $m \times 1$ vector $\mathbf{b}$ will therefore represent an underlying linear relationship between the $n$ geometry indicators and the $m$ measurement indices.

[1] Absolute-value deviations from design values of some TRC indicators (e.g. longitudinal level and alignment) and similar on-board indices are used to quantify the geometry degradation, and therefore always greater than zero. To avoid highly skewed distributions, any such quantity is represented in $Z_{q,k}$ and $Y_{q,k}$ by its (natural) logarithm.

In the case where $\boldsymbol{Y}_k$ and $\boldsymbol{Z}_k$ are both known, a multivariate regression method may be used to determine terms $\mathbf{H}$, $\mathbf{b}$, and $\mathbf{R}$ as follows. Let $\mathbf{Z}$ and $\mathbf{Y}$ be the matrices $\mathbb{R}^{N\times n}$ and matrix $\mathbb{R}^{N\times n}$ of $N$ data tuple $(\boldsymbol{Z}_k, \boldsymbol{Y}_k)$. The linear relationship between $\mathbf{Z}$ and $\mathbf{Y}$ can be expressed:

$$\mathbf{Y} = [\mathbf{Z} \quad \mathbf{1}_{N\times 1}]\begin{bmatrix}\mathbf{H}^{\mathrm{T}} \\ \mathbf{b}\end{bmatrix} + \mathbf{E} = \tilde{\mathbf{Z}}\tilde{\mathbf{H}} + \mathbf{E} \tag{5}$$

where $\mathbf{E}$ is the $\mathbb{R}^{N\times m}$ matrix of errors. The matrix $\mathbb{R}^{(n+1)\times m}$, $\tilde{\mathbf{H}}$ is obtained as following formula:

$$\tilde{\mathbf{H}} = \left(\tilde{\mathbf{Z}}^{\mathrm{T}}\tilde{\mathbf{Z}}\right)^{-1}\tilde{\mathbf{Z}}^{\mathrm{T}}\mathbf{Y} \tag{6}$$

Finally, the covariance $\mathbf{R}$ is obtained via multivariate normal distribution fitting with fixed mean $\mathbf{0}$ and data $\mathbf{E} = \left(\mathbf{Y} - \tilde{\mathbf{Z}}\tilde{\mathbf{H}}\right)$.

Since the relationship between the quantities from TRC records and the signals from the on-board system is modelled as linear, and noise distributions are assumed to be Gaussian, the Kalman filter may be employed to estimate the geometry indicators between TRC recordings $\ell$ and $\ell + 1$. For $t_\ell < \tau_{ik} \leq t_{\ell+1}$, the filtering procedure is as follows:

1. The prediction $\hat{\boldsymbol{z}}_{\ell,k+1|k}$ at time index $k+1$ is based on the estimate at time index $k$ and the degradation model in Eq. (2):

$$\hat{\boldsymbol{z}}_{\ell,k+1|k} = \begin{cases} \hat{\boldsymbol{z}}_{\ell,k|k} + \boldsymbol{\mu}\Delta t_k & k \notin \mathcal{M} \\ \boldsymbol{z}^{+} + \dfrac{1}{2}\boldsymbol{\mu}\Delta t_k & k \in \mathcal{M} \end{cases} \tag{7}$$

Similarly, the covariance $P_{k+1|k}$ is propagated using the model as follows:

$$\mathbf{P}_{k+1|k} = \begin{cases} \mathbf{P}_{k|k} + \boldsymbol{\Sigma}\Delta t_k & k \notin \mathcal{M} \\ \boldsymbol{\Sigma}_{\mathrm{z}^{+}} + \dfrac{\boldsymbol{\Sigma}\Delta t_{ik}}{2} & k \in \mathcal{M} \end{cases} \tag{8}$$

Where $\hat{\boldsymbol{z}}_{\ell,0}$ is taken as the most recent TRC measurement time ($\hat{\boldsymbol{z}}_{\ell,0} = \boldsymbol{z}_\ell$), and thus $\mathbf{P}_0 = \mathbf{0}$.

2. Update the estimate using the on-board measurement as follows:

$$\hat{\boldsymbol{z}}_{\ell,k+1|k+1} = \hat{\boldsymbol{z}}_{\ell,k+1|k} + \mathbf{K}_{k+1}\left(\boldsymbol{y}_{\ell,k+1} - \mathbf{H}\hat{\boldsymbol{z}}_{\ell,k+1|k} - \mathbf{b}\right) \tag{9}$$

and

$$\mathbf{P}_{k+1|k+1} = \mathbf{P}_{k+1|k} - \mathbf{K}_{k+1}\mathbf{S}_{k+1}\mathbf{K}_{k+1}^{T} \tag{10}$$

where matrices $\mathbf{K}_{k+1}$ and $\mathbf{S}_{k+1}$ are determined as follows:

$$\mathbf{S}_{k+1} = \mathbf{H}\mathbf{P}_{k+1|k}\mathbf{H}^{T} + \mathbf{R} \tag{11}$$

and

$$\mathbf{K}_{k+1} = \mathbf{P}_{k+1|k}\mathbf{H}^{T}\mathbf{S}_{k+1}^{-1} \tag{12}$$

Under the linear-Gaussian dynamics and measurements, the state distribution, given the measurement history, is also Gaussian with mean $\hat{\boldsymbol{z}}_{k|k}$ and covariance $\mathbf{P}_{k|k}$

In a previous work [19], the authors estimated these parameter posterior distributions via Markov Chain Monte Carlo (MCMC) sampling. The estimation thus produces a set of samples of the parameters ($\boldsymbol{\mu}$, $\mathbf{z}^+, \boldsymbol{\Sigma}_{\mathrm{z}^+}, \boldsymbol{\Sigma}$). Each sample of the joint posterior will result in a different degradation model and therefore different filtered indicators. Let $\hat{\mathbf{z}}^j_{\ell,k|k}, \mathbf{P}^j_{k|k}$ $j = 1,2, \dots, N_s$ be calculated via the Kalman Filter using the $j^{th}$ randomly drawn sample of the parameter posteriors $\left(\boldsymbol{\mu}_j, \boldsymbol{\Sigma}_j, \mathbf{z}^+_j, \boldsymbol{\Sigma}_{j,\mathrm{z}^+}\right)$. The resulting estimated state distributions $\mathcal{N}\left(\hat{\mathbf{z}}^j_{\ell,k|k}, \mathbf{P}^j_{k|k}\right)$ can be combined using a mixture of multivariate Normal distributions $\mathcal{N}\left(\hat{\mathbf{z}}^j_{\ell,k|k}, \mathbf{P}^j_{k|k}\right)$ with PDF function:

$$f_Z(\boldsymbol{Z}_{\ell,k}) = \frac{1}{N_s} \sum_{j=1}^{N_s} \frac{\exp\left(-\frac{1}{2}\left(\boldsymbol{Z}_{\ell,k} - \hat{\mathbf{z}}^j_{\ell,k|k}\right)^T \left(\mathbf{P}^j_{k|k}\right)^{-1} \left(\boldsymbol{Z}_{\ell,k} - \hat{\mathbf{z}}^j_{\ell,k|k}\right)\right)}{\sqrt{(2\pi)^n \left|\mathbf{P}^j_{k|k}\right|}} \tag{13}$$

This composite estimation therefore includes the inherent uncertainty in the degradation process as well as the parameter uncertainty induced by the estimation of the model parameters from a finite sample.

# 3 On-board Sensor system and Correlation with TRC data

## 3.1 On-board Sensor system

The case study was conducted using a Track Recording Car (TRC) within the train network of Queensland, Australia. The setup of low-cost on-board sensors followed a methodology similar to that described in previous research [8]. Prior findings have shown that sensors mounted on bogies provide the best compromise between sensitivity to track geometry and robustness to noise, compared with placements on axles or the car body [8]. As such, the configuration used in this research includes two micro-electro-mechanical systems (MEMS) accelerometers on each side (left and right) of the bogie, as shown in Figure 2. Each MEMS accelerometer is designed to capture measurements in three directions—vertical, lateral, and longitudinal—integrating these into a comprehensive data set. These accelerometers are capable of measuring a range of up to 16 g, with a noise level of $300\ \mu\mathrm{g}/\sqrt{\mathrm{Hz}}$. The sensors were programmed with a sampling rate of 2000 samples per second, and data acquisition is activated whenever the TRC is powered and operating on the network, thus capturing high-resolution data throughout its journey.

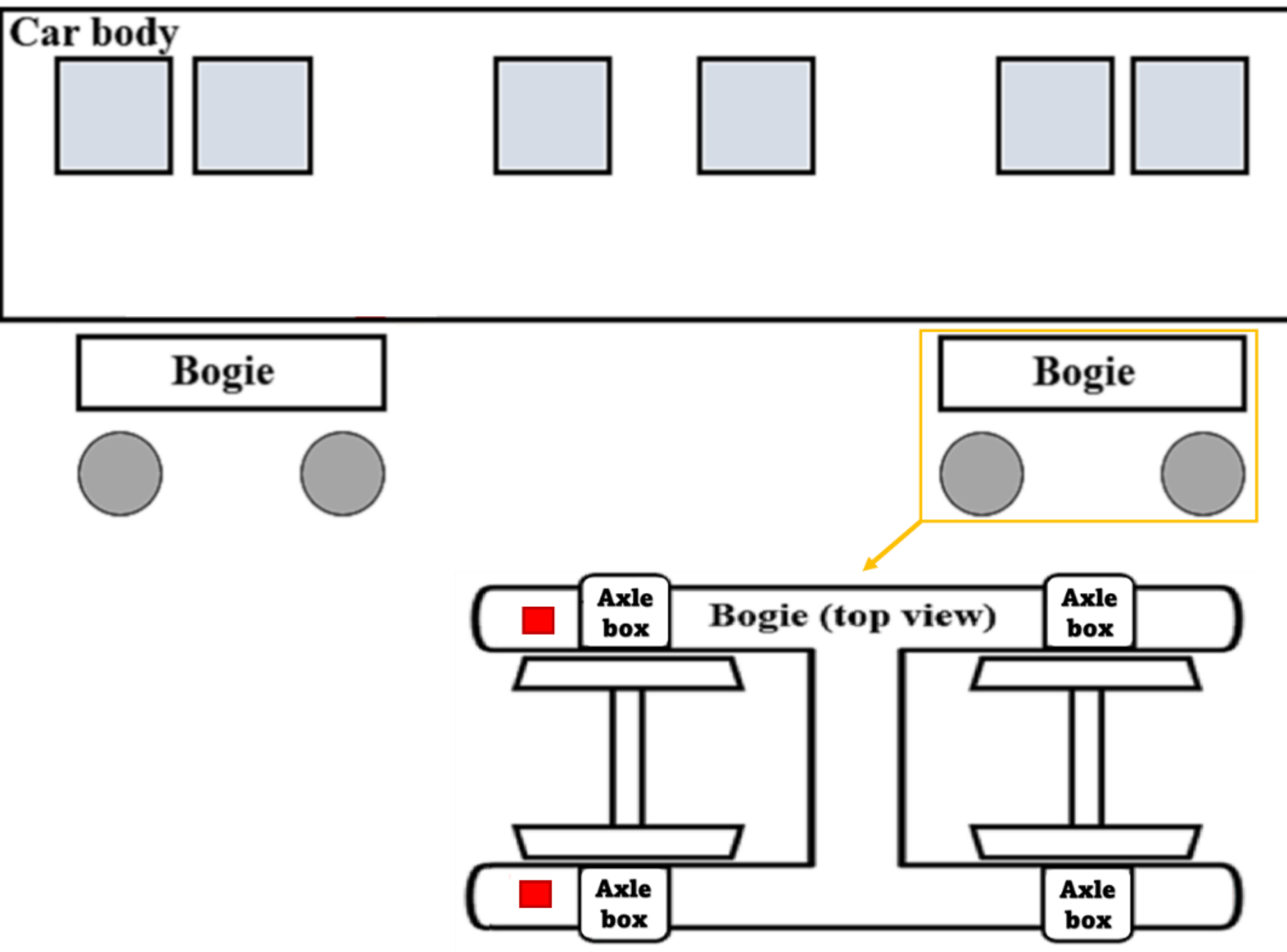

Figure 2: On-board sensors locations (approximately).

The on-board data acquisition system was built around a compact single-board Raspberry Pi computer, enhanced with an additional analog voltage input board to facilitate the collection of measurements from the two MEMS accelerometers. To ensure data integrity, signals from the power supply to the accelerometers were also recorded, allowing for the correction of any power-induced signal fluctuation that might affect the accuracy of the acceleration signals. Additionally, a GPS module was integrated into the system, providing precise location data at a rate of one sample per second. This GPS data was synchronized with the acceleration data in post-processing.

The data collection phase of this study was extensive, spanning a full year from March 2023 to March 2024. During this period, the on-board system consistently captured acceleration and GPS data as the TRC traversed the rail network. This year-long measurement campaign generated a rich data set that is critical for validating the performance of the on-board sensors and for enhancing the predictive models used in track geometry monitoring. The collected data provides a comprehensive basis for analyzing the correlation between the on-board sensor signals and the TRC measurements, ultimately contributing to more accurate and reliable track geometry predictions.

## 3.2 Signal processing and correlation with TRC data

The signal processing procedure is similar to that in the previous study [8]. Unlike that work, the left and right indicators are not considered separately in this study; instead, they are combined into a single indicator, defined as the minimum of the values from the left or right rail at the same location. The minimum value was chosen because occasional unexpected high outliers (a signal-measurement artefact) appear in one of the two rails but almost never in both rails simultaneously. On the other hand, the high correlation of left and right indicators, shown in [25], limits the loss of information inherent in this choice.

For this study, Top6 and Align10 are the longitudinal and alignment indicators, respectively. The values 6 and 10 represent the chord lengths in meters used for the computation of the geometry indices. The bandpass filter applied when converting acceleration measurements to displacement is based on the chosen minimum reference speed and chord length. For instance, the reference minimum speed for this study was chosen to be 5 m/s, and hence the data was filtered between 0.5 Hz and 40 Hz for 10m-chord quantities (e.g., Align10), and between 0.8 Hz and 40 Hz for Top6.

It should be noted that in the experimental campaign, the on-board sensors were installed on the TRC itself rather than on an in-service train, so both measurement systems operated at the same speed. In practice, speed differences between the TRC and in-service trains may affect the on-board signal characteristics, since the bandpass filtering of acceleration-to-displacement conversion depends on the vehicle speed. However, the signal processing approach adopted here uses a minimum reference speed (5 m/s in this study) to define the filter passband, which accommodates a range of operating speeds above this threshold. Any residual speed-dependent effects would manifest as additional variability in the on-board indices, which is captured by the measurement noise covariance $\mathbf{R}$ in the observation model (Eq. (3)). A systematic investigation of speed-dependent measurement error is an important direction for future work, particularly for deployment on mixed-traffic networks with varying service speeds.

To describe the relationship between track condition (as per TRC reference measurements) and onboard measurements, we start defining the quantities $\boldsymbol{Z}$ and $\boldsymbol{Y}$ , which are the logarithm of mm-measurements of actual Top6 or Align10 (TRC) and estimated Top6 or Align10 (on-board system), respectively:

$$Z_1 = \log[\mathrm{TRCTop}_6] \quad \mathrm{AND} \quad Z_2 = \log[\mathrm{TRCAlign}_{10}]$$
$$Y_1 = \log[\mathrm{OnboardTop}_6] \quad \mathrm{AND} \quad Y_2 = \log[\mathrm{OnboardAlign}_{10}]$$

Figure 3 below presents the marginal distributions of data (in normal space), collected simultaneously from the on-board sensors and the TRC for a specific track code. As illustrated in the figure, both the TRC data and the on-board sensor data exhibit a good fit with the Log-Normal distribution, indicating a consistent pattern in the underlying data across both systems. Notably, the Top indicator displays a wider range of values compared to the Alignment indicator, suggesting greater variability in this particular track geometry measure.

Figure 4 further examines these relationships by presenting the conditional distributions of TRC measurements given on-board sensor data, and vice versa. A key observation is that the conditional distribution of TRC values becomes wider as the on-board sensor indicators increase, particularly for the Top indicator. A similar behavior is observed in the conditional distribution of on-board measurements given TRC values, indicating a reciprocal dependence between the two data sources. A comparison between Top and Alignment shows that the on-board sensors capture Top-related irregularities more effectively than Alignment. This relatively lower accuracy for Alignment is consistent with findings from previous studies [8] and may be attributed to horizontal bogie motions induced by rolling effects or dynamic impulsive events.

Despite these similarities in overall trends, the TRC and on-board measurements are not identical. The conditional distributions indicate that on-board sensor readings tend to be lower than the corresponding TRC values, reflecting systematic differences between the two measurement systems. These discrepancies may arise from differences in sensor quality, calibration, installation location, or operating conditions. Nevertheless, when combined with TRC data, the high-frequency on-board measurements provide valuable supplementary information that can enhance the accuracy and reliability of track geometry estimation and prediction.

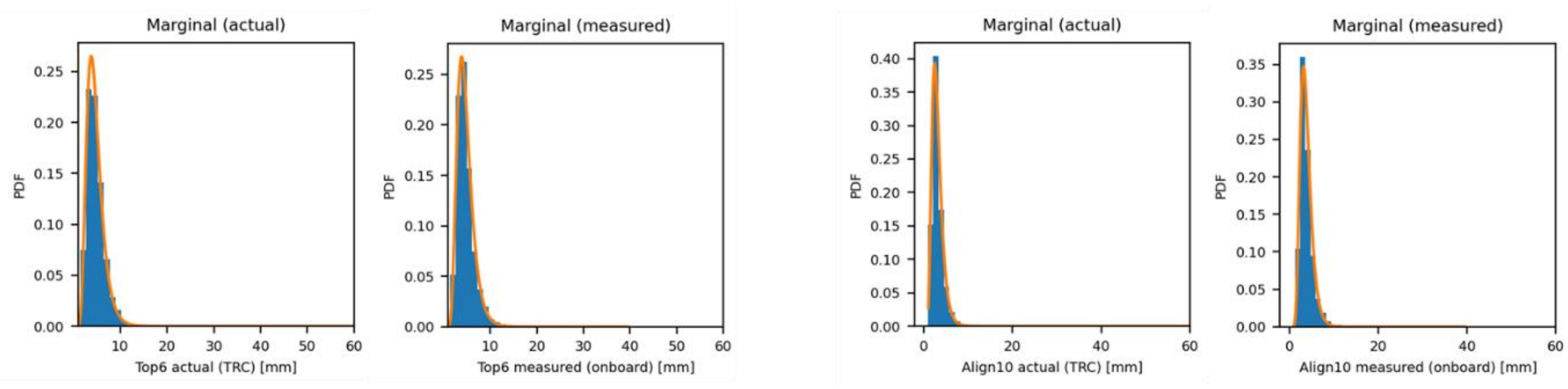

Figure 3: Illustration of Marginal distributions of TRC and on-board data

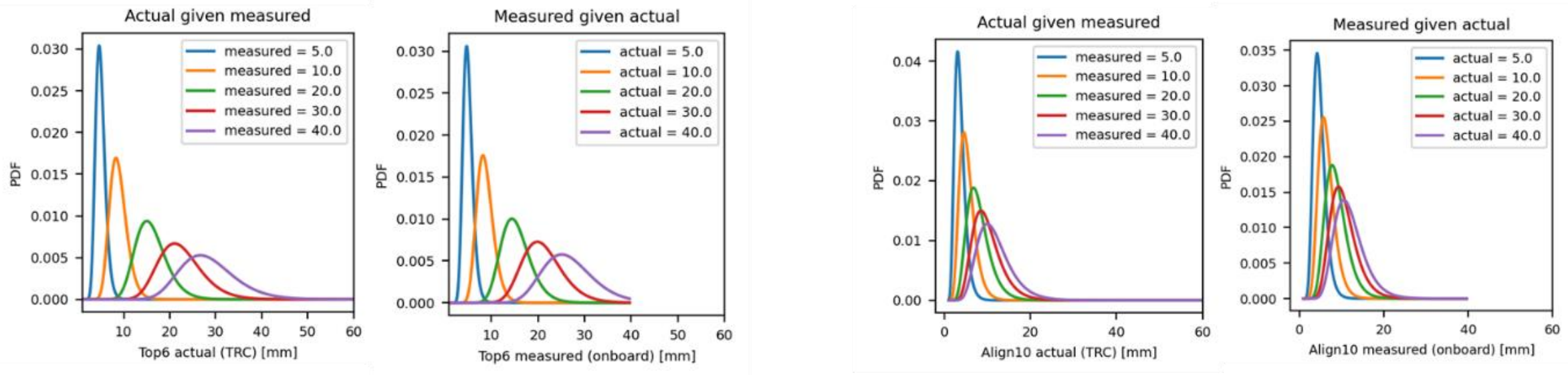

Figure 4: Illustration of Conditional distributions of TRC and on-board data

Because the TRC data and on-board signals were collected concurrently in this experimental campaign, the parameters of the relation function between on-board signal $\boldsymbol{Y}_k$ and track geometry indicators from TRC $\boldsymbol{Z}_k$, i.e. $\mathbf{H}$, $\mathbf{b}$ and $\mathbf{R}$ were obtained via multivariate regression approaches as described in Equations (5) and (6) above.

The logarithmic transformation of the considered track geometry indicators is motivated by both physical and statistical considerations. Track geometry indicators such as Top and Alignment represent absolute-value deviations from design values, which are inherently non-negative. Modelling in log-space naturally enforces this non-negativity constraint without requiring truncation or bounded distributions. In addition, the marginal distributions of both TRC and on-board sensor data are well-approximated by log-normal distributions, as shown in Figure 3, which implies that a logarithmic transformation yields approximately normally distributed quantities. Furthermore, the conditional distributions in Figure 4 show increasing dispersion with increasing indicator magnitude in normal space, a heteroscedastic pattern that is substantially reduced in log-space, improving the validity of the constant-covariance assumption in the observation model (Eq.(3)). Our previous study [19] confirmed that modelling accuracy in log-space is comparable to that achieved in normal space, further supporting this choice.

# 4 Sensor fusion for estimation of track geometry with Kalman filter

## 4.1 Validation of Kalman filter prediction track geometry indicators

To validate the use of the Kalman filter for predicting track geometry, a comprehensive test was conducted with the TRC of our industry partner. A low-cost on-board system, capable of transmitting signals in real-time via mobile networks during TRC operations, was installed on the TRC. This setup allowed for the simultaneous collection of data from the TRC and signals from the on-board system, enabling a detailed analysis of their correlations and an assessment of the Kalman filter's effectiveness.

For validation purposes, data from tracks that the TRC, equipped with the on-board sensor system, visited multiple times were selected for analysis. The track geometry predictions were carried out using a combination of the degradation model and the Kalman filter, which incorporated the real-time signals from the on-board sensors. The corresponding TRC measurements were then employed to validate these predictions, ensuring that the Kalman filter accurately reflected the actual track conditions. This validation process not only demonstrated the robustness of the Kalman filter but also highlighted the value of integrating real-time sensor data into predictive models for more reliable track geometry estimation. The data availability for validation is described in the following table:

Table 1: Data availability for validation

| TIME | WEEK 0 | WEEK 11 | WEEK 19 | WEEK 27 | WEEK 34 |
| --- | --- | --- | --- | --- | --- |

| TRC AVAILABILITY | Yes | Yes | Yes | Yes | Yes |
|---|---|---|---|---|---|
| SENSOR AVAILABILITY | No | Yes | Yes | Yes | Yes |

In this study, two key track geometry indicators were considered for prediction: the longitudinal indicator known as Top and the alignment indicator known as Alignment. The prediction models, along with the actual data from the TRC and the signals from the on-board sensors, are depicted in Figure 5. While the on-board signals cannot be directly used to predict the actual track geometry data due to discrepancies in their values, their overall patterns are strikingly similar to those of the actual indicators. This similarity allows the on-board signals to be effectively utilized in a predictive model.

Despite the limited number of data points available over an extended time horizon, the use of the Kalman filter demonstrates significant benefits in reducing the width of the credibility zone, thereby enhancing prediction confidence. Moreover, the Kalman filter's ability to incorporate real-time updates from the on-board signals ensures that the prediction patterns closely follow those of the signals. As a result, the predictions align closely with the actual track geometry indicators, providing a more accurate and reliable estimation. This approach underscores the practical advantage of integrating on-board sensor data with predictive models, particularly in scenarios where direct measurement is challenging or infrequent.

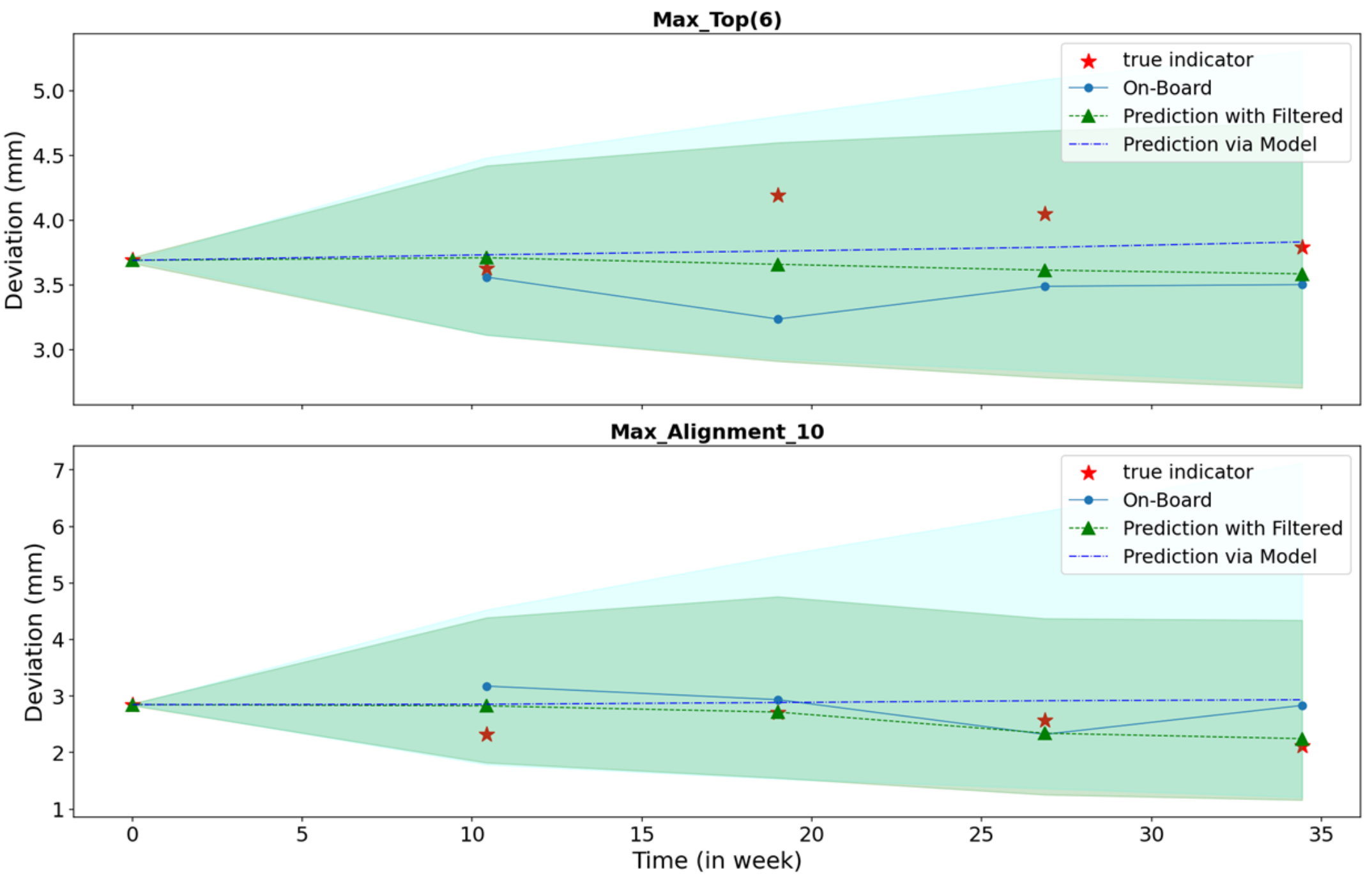

Figure 5: Validation of Prediction with filter vs. actual data.

## 4.2 Effects of measurement frequency on track geometry prediction credibility

Following the testing phase of the developed sensors on the Track Recording Car (TRC), these sensors will be installed on service trains to frequently record signals related to track geometry indicators. A key concern for rail operators is determining the optimal number of sensor systems required for a given track or network. More specifically, the question arises: how frequently should signals be recorded to achieve an optimal estimation of track geometry while balancing the capital costs? This section presents an analysis of the relationship between measurement frequency and the credibility of track geometry predictions. The analysis is conducted using Monte Carlo simulations based on the relationship between sensor signals and TRC records discussed earlier in Section 3.2. The simulation procedure is as follows:

Initialization: select a segment and initial TRC record. Set the time horizon $t_0, t_1, \dots, t_K$ and measurement interval $\Delta t = t_{k+1} - t_k$, where $t_0$ is the time of the TRC record.

Step 1: Generate a random geometry indicator path from $t_1$ to $t_K$, $\boldsymbol{z}_1, \boldsymbol{z}_2, \dots, \boldsymbol{z}_K$, by using the degradation model.

Step 2: Generate the signal from the on-board sensor $\boldsymbol{y}_k$ according to each $\boldsymbol{z}_k, k = 1..K$.

Step 3: Determine the predicted values of indicators $\hat{q}_k$ according to $o_k$ and the degradation model.

Step 4: Calculate the width $W_k$ of prediction credibility zone via formula [45,46]:

$$W_k = \left|P_{k|k}\right|^{\frac{1}{n}} \tag{14}$$

Where $n$ is the number of indicators.

Figure 6 below illustrates a simulated track geometry path over one year (i.e., 52 weeks), along with predictions of Top and Alignment using different sensor measurement intervals. The simulation demonstrates how on-board sensor data can be used in conjunction with a Kalman filter to enhance track geometry predictions. As observed in the validation process, while the signals from the on-train sensors exhibit noise and deviations from the actual simulated indicator values, the overall signal patterns closely follow the trends of the real track geometry indicators. This adaptability of the sensor signals, despite the disturbances, makes them valuable for improving the prediction accuracy of the track geometry indicators.

The figure clearly shows that predictions made using the Kalman filter result in significantly smaller credibility zones (depicted in green) compared to predictions based solely on the degradation model (depicted in cyan). This indicates a higher level of confidence in the predictions when the Kalman filter is applied, as the filter effectively integrates the sensor data to reduce uncertainty. Additionally, unlike the predictions made with the degradation model alone, the credibility zone associated with the Kalman filter remains stable over time.

The stable credibility zone implies a practical advantage for rail operators: in case regulations and operational constraints permit, the on-board system can support more informed and flexible use of the TRC while also enabling prompt detection of geometry deviations. Yet, a very thorough empirical characterization of the measurement system performance and error is also required before its practical use in order to ensure that the measurement-error model used in such optimization is sufficiently representative of its real behavior, in particular highlighting whether particular effects could change its performance over time or under different operating environments.

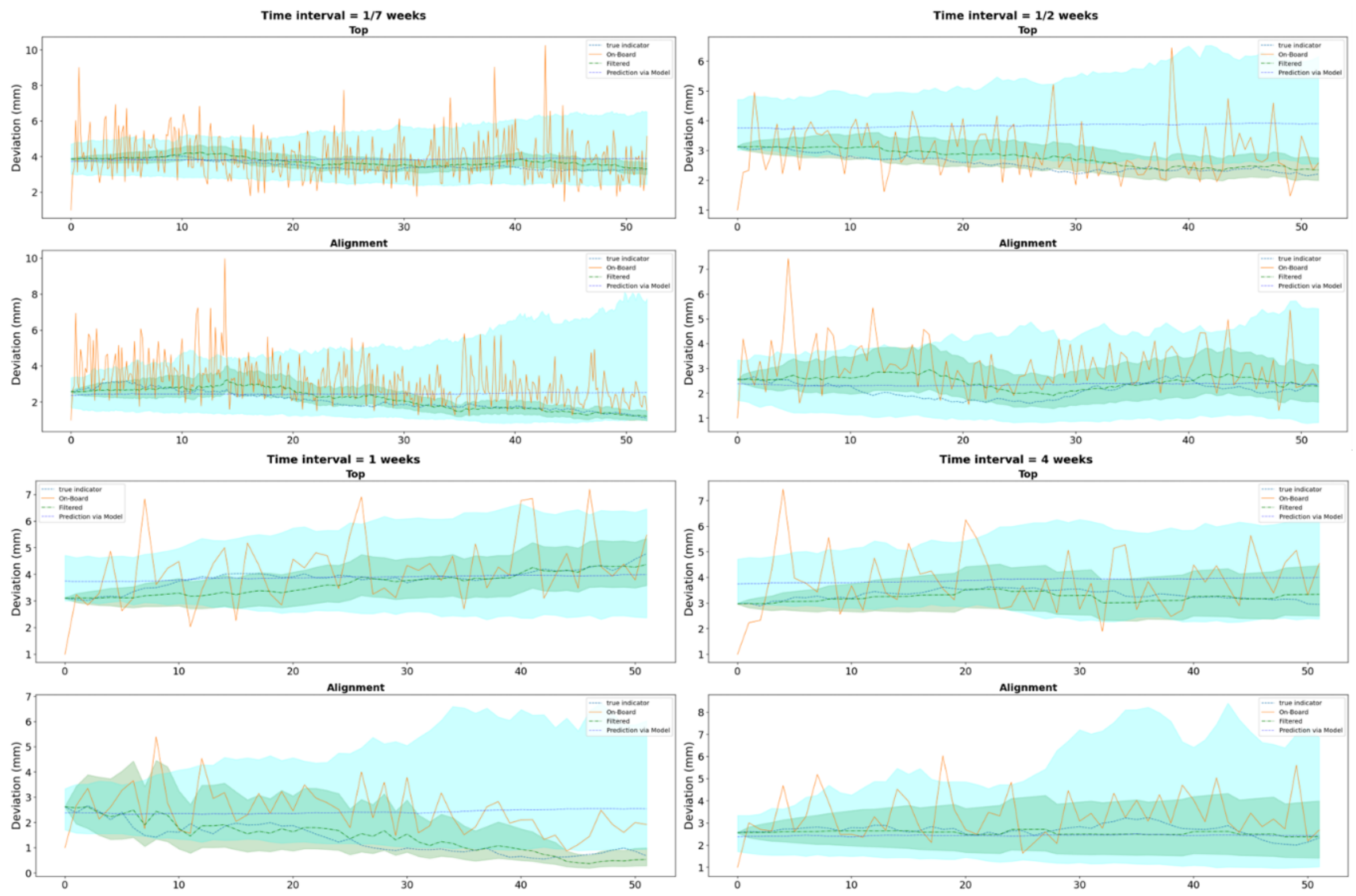

Figure 6: Prediction with Kalman filter in different sensor measurement intervals.

Figure 7 shows the relative width of credibility zone $W_k$ calculated via Eq. $(14)$ for different sensor measurement intervals and Figure 8 illustrates these widths of credibility zones with the assumption that the means of Top and Alignment are maintained at 12mm and 10mm, respectively. These zones are also compared to the credibility zone derived from the degradation model alone, serving as a benchmark. Initially, the widths of the prediction credibility zones are relatively narrow, reflecting high confidence in the early stages. As time progresses, the widths increase due to accumulating uncertainty, eventually stabilizing at a consistent level.

A key observation is that the stabilization of the credibility zone occurs more quickly with shorter measurement intervals. In other words, the more frequently sensor data is collected, the sooner the predictions achieve a steady and reliable level of confidence. Notably, predictions incorporating on-board sensors show a significantly smaller credibility zone after approximately two months, even when the measurement interval extends up to eight weeks. This demonstrates the substantial benefits of integrating on-train sensors into the track geometry estimation process.

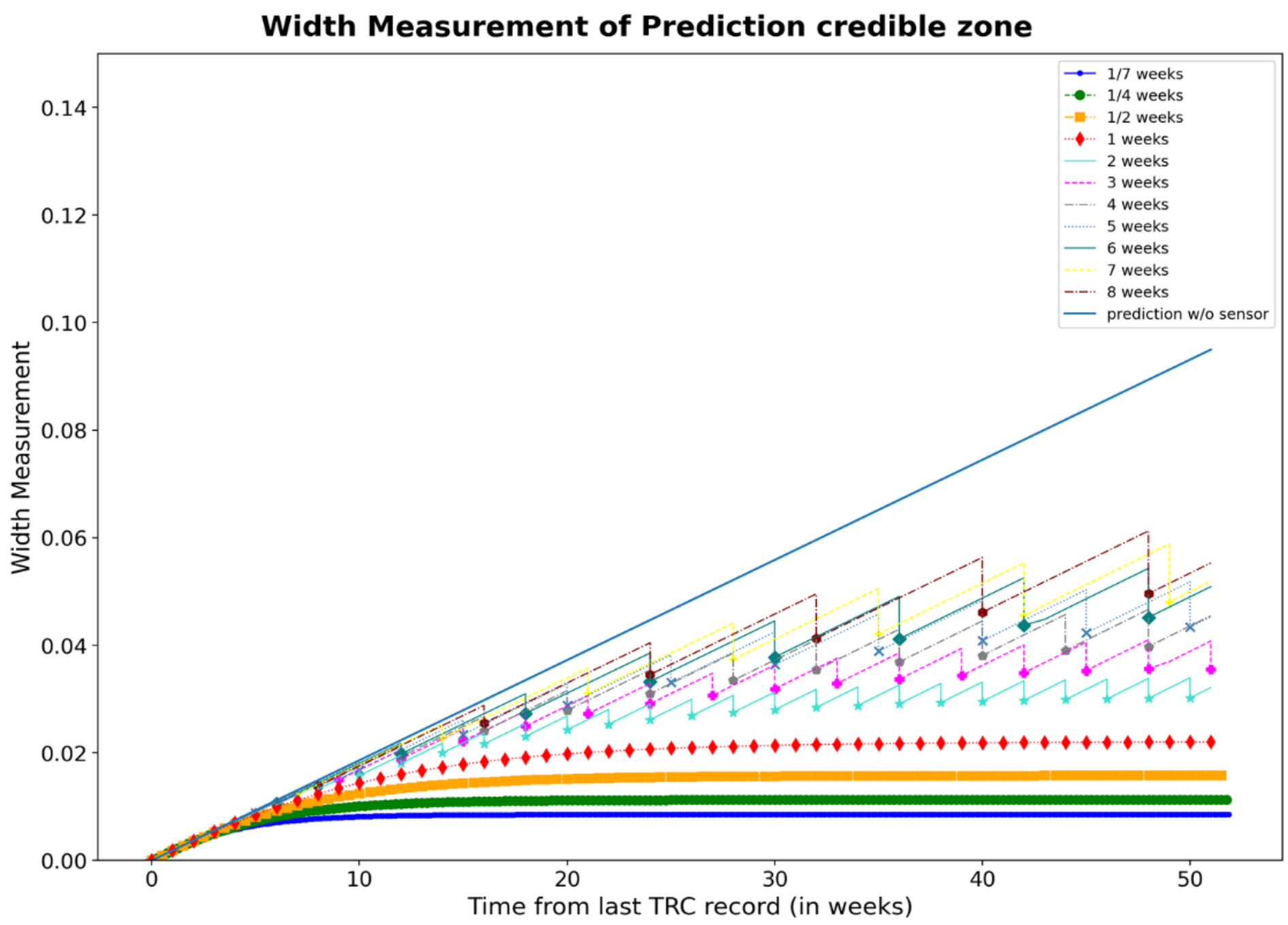

Figure 7: Widths of credibility zone with different sensor measurement intervals.

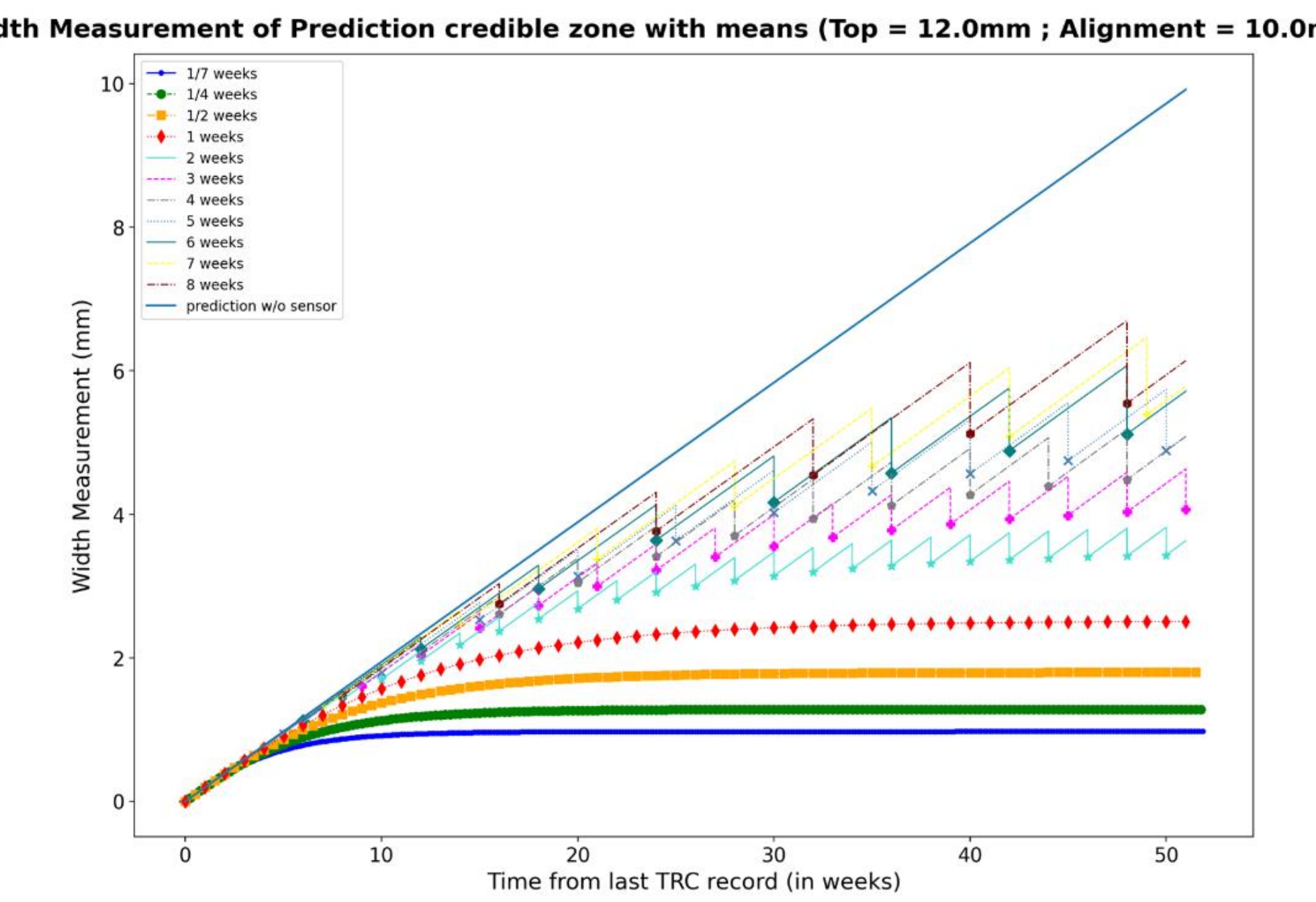

Figure 8: Illustration of credibility zone width for specific case of Top = 12mm and Alignment = 10mm.

Table 2 summarizes the widths of the credibility intervals for track geometry predictions at 3- and 6-month horizons across fast, typical, and slow degradation scenarios, considering different on-board sensor measurement intervals and a benchmark case without on-board data. The results clearly show that incorporating low-cost on-board sensor measurements leads to substantial reductions in predictive uncertainty compared with using TRC data alone.

At the 3-month prediction horizon, the inclusion of weekly on-board measurements reduces the credibility interval width by approximately 55–70% for Top indicators and 20–50% for Alignment indicators, depending on the degradation rate. Even when the measurement interval is extended to four weeks, the reduction in uncertainty remains significant, typically on the order of 40–60% for Top and 15–40% for Alignment. These improvements are most pronounced under fast degradation scenarios, where predictions without on-board data exhibit the largest uncertainty growth.

At the 6-month horizon, the benefits of on-board sensing become even more evident. For fast degradation cases, weekly on-board measurements reduce the credibility interval width by up to 80–90% for Top and 40–60% for Alignment compared with the no–on-board baseline. For typical and slow degradation scenarios, the reductions remain substantial, generally ranging from 50–75% for Top and 30–55% for Alignment. Although longer measurement intervals lead to wider credibility zones, the results indicate that even sparse on-board data (e.g., four-week intervals) can still reduce long-term predictive uncertainty by more than 40% in many cases.

Overall, these reductions in credibility interval width demonstrate the effectiveness of the proposed hybrid framework in constraining uncertainty growth over time. By fusing infrequent high-accuracy TRC measurements with frequent but low-cost on-board sensor data, the approach substantially improves the robustness and reliability of long-term track geometry predictions. This capability is particularly valuable for sections experiencing rapid degradation, where improved uncertainty control can directly support more proactive maintenance planning and more efficient allocation of inspection resources.

Table 2: Widths of credibility zone (in log-space) of prediction at 3 and 6 months

| | | **Fast Degradation** | | **Typical Degradation** | | **Slow Degradation** | |
|---|---|---|---|---|---|---|---|
| | | *Top* | *Alignment* | *Top* | *Alignment* | *Top* | *Alignment* |
| C.I at 3 months | No on-board | [0.540,1.656] | [0.033,0.193] | [0.023,0.105] | [0.204,0.780] | [0.028,0.134] | [0.044,0.222] |
| | 1 week | [0.184,0.360] | [0.031,0.147] | [0.020,0.067] | [0.154,0.386] | [0.024,0.079] | [0.041,0.164] |
| | 2 weeks | [0.229,0.423] | [0.032,0.155] | [0.021,0.078] | [0.171,0.472] | [0.025,0.094] | [0.042,0.184] |
| | 4 weeks | [0.274,0.494] | [0.032,0.163] | [0.022,0.086] | [0.182,0.545] | [0.026,0.105] | [0.043,0.196] |
| C. | No on-board | [0.866,3.252] | [0.067,0.594] | [0.046,0.211] | [0.409,1.560] | [0.056,0.268] | [0.088,0.444] |

| 1 week | [0.184,0.379] | [0.054,0.335] | [0.030,0.075] | [0.191,0.400] | [0.034,0.085] | [0.067,0.200] |
|---|---|---|---|---|---|---|
| 2 weeks | [0.275,0.564] | [0.059,0.398] | [0.036,0.105] | [0.255,0.580] | [0.042,0.122] | [0.076,0.269] |
| 4 weeks | [0.334,0.623] | [0.062,0.475] | [0.040,0.129] | [0.299,0.723] | [0.047,0.151] | [0.081,0.318] |

# 5 Conclusion

This study proposes a collaborative approach to predicting multiple track geometry indicators by integrating a degradation model based on TRC data with data from sensors installed on in-service trains. This approach is implemented through a Kalman filter, applied to the natural logarithm of the track geometry indicators.

To ensure scientific validation for the proposed methodology, a special campaign was conducted with the on-board measurement system installed directly on a TRC, rather than on an in-service train. In particular, the data from the on-board measurement system was used in the predictive model together with only a portion of the available TRC reference measurements, while leaving the rest for the validation of the predictions. The measurements were performed within the train network of Queensland, Australia, with data collected over one year.

An analysis was also conducted to assess the impact of measurement frequency on the predictive credibility zone. The results suggest that reducing the measurement interval to one week or less significantly reduces the width of the credibility zone and maintains its stability over time, even as the prediction horizon extends. This frequency seems reasonable for most tracks, assuming at least one regular train service per track is available, with rolling stock equipped for on-board track geometry estimation.

Future work will focus on the development of TRC inspection planning and tamping decision strategies for the entire track network, leveraging the collaborative predictions from both the degradation model and the on-board sensors.

# Acknowledgments

This study has been undertaken as part of the Australian Research Council (ARC) Linkage Project LP200100382.

During the preparation of this work the authors used Claude (Anthropic) in order to support writing refinement, including improving grammar, enhancing conciseness, and clarifying the text. After using this tool/service, the authors reviewed and edited the content as needed and take(s) full responsibility for the content of the published article.

## References

[1] Arcieri G, Hoelzl C, Schwery O, Straub D, Papakonstantinou KG, Chatzi E. Bridging POMDPs and Bayesian decision making for robust maintenance planning under model uncertainty: An application to railway systems. Reliability Engineering & System Safety 2023;239:109496. https://doi.org/10.1016/j.ress.2023.109496.

[2] Lee JS, Yeo I-H, Bae Y. A stochastic track maintenance scheduling model based on deep reinforcement learning approaches. Reliability Engineering & System Safety 2024;241:109709. https://doi.org/10.1016/j.ress.2023.109709.

[3] Wu Q, Azad AK, Cole C, Spiryagin M. Identify severe track geometry defect combinations for maintenance planning. International Journal of Rail Transportation 2022;10:95–113.

[4] Khajehei H, Ahmadi A, Soleimanmeigouni I, Nissen A. Allocation of effective maintenance limit for railway track geometry. Structure and Infrastructure Engineering 2019:1–16. https://doi.org/10.1080/15732479.2019.1629464.

[5] Saleh A, Remenyte-Prescott R, Prescott D, Chiachío M. Intelligent and adaptive asset management model for railway sections using the iPN method. Reliability Engineering & System Safety 2024;241:109687. https://doi.org/10.1016/j.ress.2023.109687.

[6] Soleimanmeigouni I, Ahmadi A, Kumar U. Track geometry degradation and maintenance modelling: A review. Proceedings of the Institution of Mechanical Engineers, Part F: Journal of Rail and Rapid Transit 2018;232:73–102.

[7] Wang Y, Wang P, Wang X, Liu X. Position synchronization for track geometry inspection data via big-data fusion and incremental learning. Transportation Research Part C: Emerging Technologies 2018;93:544–65. https://doi.org/10.1016/j.trc.2018.06.018.

[8] Zhang H, Chin ZY, Borghesani P, Pitt J, Cholette ME. Evaluation of onboard sensors for track geometry monitoring against conventional track recording measurements. Measurement 2024;229:114354. https://doi.org/10.1016/j.measurement.2024.114354.

[9] A. Peinado Gonzalo, R. Horridge, H. Steele, E. Stewart, M. Entezami. Review of Data Analytics for Condition Monitoring of Railway Track Geometry. IEEE Transactions on Intelligent Transportation Systems 2022;23:22737–54. https://doi.org/10.1109/TITS.2022.3214121.

[10] Rebello S, Cholette ME, Truong-Ba H, Reddy V, Rosser A, Watkin T. Railway Track Geometry Degradation Modelling and Prediction for Maintenance Decision Support. In: Pinto JOP, Kimpara MLM, Reis RR, Seecharan T, Upadhyaya BR, Amadi-Echendu J, editors. 15th WCEAM Proceedings, Cham: Springer International Publishing; 2022, p. 422–32. https://doi.org/10.1007/978-3-030-96794-9_39.

[11] Letot C, Soleimanmeigouni I, Ahmadi A, Dehombreux P. An adaptive opportunistic maintenance model based on railway track condition prediction. IFAC-PapersOnLine 2016;49:120–5. https://doi.org/10.1016/j.ifacol.2016.11.021.

[12] Mishra M, Odelius J, Thaduri A, Nissen A, Rantatalo M. Particle filter-based prognostic approach for railway track geometry. Mechanical Systems and Signal Processing 2017;96:226–38.

[13] Sharma S, Cui Y, He Q, Mohammadi R, Li Z. Data-driven optimization of railway maintenance for track geometry. Transportation Research Part C: Emerging Technologies 2018;90:34–58. https://doi.org/10.1016/j.trc.2018.02.019.

[14] Soleimanmeigouni I, Xiao X, Ahmadi A, Xie M, Nissen A, Kumar U. Modelling the evolution of ballasted railway track geometry by a two-level piecewise model. Structure and Infrastructure Engineering 2018;14:33–45. https://doi.org/10.1080/15732479.2017.1326946.

[15] Vale C, M. Lurdes S. Stochastic model for the geometrical rail track degradation process in the Portuguese railway Northern Line. Reliability Engineering & System Safety 2013;116:91–8. https://doi.org/10.1016/j.ress.2013.02.010.
[16] Mercier S, Meier-Hirmer C, Roussignol M. Bivariate Gamma wear processes for track geometry modelling, with application to intervention scheduling. Structure and Infrastructure Engineering 2012;8:357–66.
[17] Soleimanmeigouni I, Ahmadi A, Nissen A, Xiao X. Prediction of railway track geometry defects: a case study. Structure and Infrastructure Engineering 2020;16:987–1001. https://doi.org/10.1080/15732479.2019.1679193.
[18] Lasisi A, Attoh-Okine N. Principal components analysis and track quality index: A machine learning approach. Transportation Research Part C: Emerging Technologies 2018;91:230–48. https://doi.org/10.1016/j.trc.2018.04.001.
[19] Truong-Ba H, Rebello S, Cholette ME, Reddy V, Borghesani P. Bayesian multivariate track geometry degradation modeling and its use in condition-based inspection. Railway Engineering Science 2025. https://doi.org/10.1007/s40534-025-00394-4.
[20] Li C, He Q, Wang P. Estimation of railway track longitudinal irregularity using vehicle response with information compression and Bayesian deep learning. Computer-Aided Civil and Infrastructure Engineering 2022;37:1260–76. https://doi.org/10.1111/mice.12802.
[21] Weston P, Roberts C, Yeo G, Stewart E. Perspectives on railway track geometry condition monitoring from in-service railway vehicles. Vehicle System Dynamics 2015;53:1063–91.
[22] La Paglia I, Carnevale M, Corradi R, Di Gialleonardo E, Facchinetti A, Lisi S. Condition monitoring of vertical track alignment by bogie acceleration measurements on commercial high-speed vehicles. Mechanical Systems and Signal Processing 2023;186:109869. https://doi.org/10.1016/j.ymssp.2022.109869.
[23] Tsunashima H, Hirose R. Condition monitoring of railway track from car-body vibration using time–frequency analysis. Vehicle System Dynamics 2022;60:1170–87. https://doi.org/10.1080/00423114.2020.1850808.
[24] Lederman G, Chen S, Garrett J, Kovačević J, Noh HY, Bielak J. Track-monitoring from the dynamic response of an operational train. Mechanical Systems and Signal Processing 2017;87:1–16. https://doi.org/10.1016/j.ymssp.2016.06.041.
[25] Pereira P, Ribeiro F, Motta R, Neto AG, Bernucci L, Júnior EF, et al. Perspectives of assessing the geometric condition of railway tracks using a device for measuring the topographic profile. Measurement 2025;249:117020. https://doi.org/10.1016/j.measurement.2025.117020.
[26] Circelli M, Kaviani N, Licciardello R, Ricci S, Rizzetto L, Arabani SS, et al. Track geometry monitoring by an on-board computer-vision-based sensor system. Transportation Research Procedia 2023;69:257–64. https://doi.org/10.1016/j.trpro.2023.02.170.
[27] B. Olivier, F. Guo, Y. Qian, D. P. Connolly. A Review of Computer Vision for Railways. IEEE Transactions on Intelligent Transportation Systems 2025;26:11034–65. https://doi.org/10.1109/TITS.2025.3552011.
[28] Muñoz S, Ros J, Urda P, Escalona JL. Estimation of lateral track irregularity through Kalman filtering techniques. IEEE Access 2021;9:60010–25.
[29] Muñoz S, Urda P, Escalona JL. Experimental measurement of track irregularities using a scaled track recording vehicle and Kalman filtering techniques. Mechanical Systems and Signal Processing 2022;169:108625.
[30] De Rosa A, Alfi S, Bruni S. Estimation of lateral and cross alignment in a railway track based on vehicle dynamics measurements. Mechanical Systems and Signal Processing 2019;116:606–23. https://doi.org/10.1016/j.ymssp.2018.06.041.
[31] Lee JS, Choi IY, Kim S, Moon DS. Kinematic modeling of a track geometry using an unscented Kalman filter. Measurement 2016;94:707–16.

[32] Xiao X, Xu X-Y, Zhu Q, Ren W-X. A Bayesian Kalman filter algorithm for quantifying estimation uncertainty of track irregularity on bridges with randomness in system parameters. Mechanical Systems and Signal Processing 2024;209:111097.
[33] Hoelzl C, Dertimanis V, Landgraf M, Ancu L, Zurkirchen M, Chatzi E. Chapter 9 - On-board monitoring for smart assessment of railway infrastructure: A systematic review. In: Alavi AH, Feng MQ, Jiao P, Sharif-Khodaei Z, editors. The Rise of Smart Cities, Butterworth-Heinemann; 2022, p. 223–59. https://doi.org/10.1016/B978-0-12-817784-6.00015-1.
[34] Paixão A, Fortunato E, Calçada R. Smartphone's Sensing Capabilities for On-Board Railway Track Monitoring: Structural Performance and Geometrical Degradation Assessment. Advances in Civil Engineering 2019;2019:1729153. https://doi.org/10.1155/2019/1729153.
[35] Koren Y, Rendle S, Bell R. Advances in Collaborative Filtering. In: Ricci F, Rokach L, Shapira B, editors. Recommender Systems Handbook, New York, NY: Springer US; 2022, p. 91–142. https://doi.org/10.1007/978-1-0716-2197-4_3.
[36] Khodarahmi M, Maihami V. A Review on Kalman Filter Models. Archives of Computational Methods in Engineering 2023;30:727–47. https://doi.org/10.1007/s11831-022-09815-7.
[37] Xie L, Soh YC. Robust Kalman filtering for uncertain discrete-time systems. IEEE Transactions on Automatic Control 1994;39:1310–4.
[38] Gao B, Gao S, Zhong Y, Hu G, Gu C. Interacting multiple model estimation-based adaptive robust unscented Kalman filter. International Journal of Control, Automation and Systems 2017;15:2013–25.
[39] Arasaratnam I, Haykin S. Cubature kalman filters. IEEE Transactions on Automatic Control 2009;54:1254–69.
[40] Farahi F, Yazdi HS. Probabilistic Kalman filter for moving object tracking. Signal Processing: Image Communication 2020;82:115751.
[41] Lu P. Nonlinear predictive controllers for continuous systems. Journal of Guidance, Control, and Dynamics 1994;17:553–60.
[42] Calvet LE, Czellar V, Ronchetti E. Robust filtering. Journal of the American Statistical Association 2015;110:1591–606.
[43] Jouin M, Gouriveau R, Hissel D, Péra M-C, Zerhouni N. Particle filter-based prognostics: Review, discussion and perspectives. Mechanical Systems and Signal Processing 2016;72–73:2–31. https://doi.org/10.1016/j.ymssp.2015.11.008.
[44] F. Gustafsson. Particle filter theory and practice with positioning applications. IEEE Aerospace and Electronic Systems Magazine 2010;25:53–82. https://doi.org/10.1109/MAES.2010.5546308.
[45] Salibián-Barrera M, Van Aelst S, Willems G. Principal Components Analysis Based on Multivariate MM Estimators With Fast and Robust Bootstrap. Journal of the American Statistical Association 2006;101:1198–211. https://doi.org/10.1198/016214506000000096.
[46] Taskinen S, Croux C, Kankainen A, Ollila E, Oja H. Influence functions and efficiencies of the canonical correlation and vector estimates based on scatter and shape matrices. Journal of Multivariate Analysis 2006;97:359–84. https://doi.org/10.1016/j.jmva.2005.03.005.